\documentclass{article}

\usepackage{arxiv}

\usepackage[utf8]{inputenc} 
\usepackage[T1]{fontenc}    
\usepackage{hyperref}       
\usepackage{url}            
\usepackage{booktabs}       
\usepackage{amsfonts}       
\usepackage{nicefrac}       
\usepackage{microtype}      
\usepackage{lipsum}
\usepackage{graphicx}
\graphicspath{ {./images/} }

\usepackage{subcaption}

\title{Robust Multi-Domain Mitosis Detection}

\author{
 Mustaffa Hussain \\
  Onward Assist\\
  Hyderabad, 500032 \\
  \texttt{mustaffa@onwardhealth.co} \\
   \And
 Ritesh Gangnani \\
  Onward Assist\\
  Hyderabad, 500032 \\
  \texttt{ritesh@onwardhealth.co} \\
  \And
Sasidhar kadiyala \\
  Onward Assist\\
  Hyderabad, 500032 \\
  \texttt{sasidhar.kadiyala1@gmail.com} \\
}

\begin{document}
\maketitle
\begin{abstract}
Domain variability is a common bottle neck in developing generalisable algorithms for various medical applications. Motivated by the observation that the domain variability of the medical images is to some extent compact, we propose to learn a target representative feature space through unpaired image to image translation (CycleGAN). We comprehensively evaluate the performance and usefulness by utilising the transformation to mitosis detection with candidate proposal and classification. This work presents a simple yet effective multi-step mitotic figure detection algorithm developed as a baseline for the MIDOG challenge. On the preliminary test set, the
algorithm scores an F1 score of 0.52.

\end{abstract}


\section{Introduction}
Staining is a standard procedure in histopathology. It aids the pathologists in diagnosis by highlighting the morphological structures of interest under examination in microscopes. Differences in staining are common across lab. It can occur from tissue fixation and processing, staining protocols, and section thickness, etc. So domain-shift is inevitable to some degree even for the same highly standardized laboratory due to differences in tissue handling and manual steps in specimen preparation; however, is especially notable between different laboratories. The past decade has seen the rise of computer-aided diagnostics and resulted in many vendors producing digital scanners. Digitization of tissue into whole slide images (WSIs) is the first step to diagnostics. The use of different scanners introduces new sources of variation, such as type to the scanner, scanning resolutions, differences in digital post-processing algorithms, and storage format, etc. The most important source of a domain shift is the whole slide image acquisition due to highly variable color representation and other image parameters between different types of whole slide scanners.  

\section{Material and methodology}
In this section, we discuss data and implementation details.

The reference algorithm was developed on the official training subset of the MIDOG dataset \cite{midog_data}. We did not use any additional datasets and had no access to the preliminary test during method development. The algorithm is based on a publicly available implementation of CycleGAN  \cite{cyclegan} and RetinaNet \cite{retina_net} with a classification block of EfficientNets \cite{efficientnet} to improve detection.

\subsection{Dataset} 
The MIDOG training subset consists of 200 Whole Slide Images (WSIs) from human breast cancer tissue samples stained with routine Hematoxylin \& Eosin (H\&E) dye. The samples were digitized with four slide scanning systems: the Hamamatsu XR NanoZoomer 2.0, the Hamamatsu S360, the Aperio ScanScope CS2, and the Leica GT450, resulting
in 50 WSIs per scanner. For the slides of three scanners, a selected field of interest sized approximately 2 mm2 (equivalent to ten high power fields) was annotated for mitotic figures and hard negative look-alikes. These annotations were collected in a multi-expert blinded set-up. For the Leica GT450, no annotations were available. 

\subsection{Methodology}
We tackle the problem in 3 stages namely stain generalization using CycleGAN, candidate detection using RetinaNet-R-50-FPN-3x, and classification using EfficientNet-B6. More info about CycleGAN can be found here \cite{cyclegan}. We use patches of 1024x1024 from Hamamatsu XR NanoZoomer 2.0, Hamamatsu S360, Aperio ScanScope CS2 as domain A and 1024x1024 patches from Leica GT450 as domain B. The unpaired image to image mapping is learned from domain A to domain B as the target domain. CycleGAN was trained for 200 epochs with standard cycle consistency loss and a learning rate of 0.0002. Figure \ref{fig:fig_cyc} shows original 1024x1024 patches and their corresponding stain generalized patches at 1024x1024 using CycleGAN. Following stain generalization, object detection training was done for the candidate proposal. For this 150 WSIs were used with a 7:3 train-test split where an 8:2 split was used from train data for train and validation split. Patches of 512x512 were used in the training of RetinaNet-R-50-FPN-3x \cite{wu2019detectron2} with non maximum suppression(NMS) threshold of 0.1, classification score threshold of 0.35 for 10000 epochs with a learning rate of 0.0001. The EfficientNet-B6 classification network was trained on the bounding boxes annotations provided for 150 WSIs with a 7:3train-test split where 8:2 split was used from train data for train and validation split. All bounding box annotations were fed in a standard 50x50 patch size. Offline augmentations such as vertical flips, random rotations, etc were used for classification data generation. Online augmentations like height shift, width shift, horizontal flip, and vertical flips were used. The classification network was trained for 50 epochs with early stopping with a learning rate of 0.001. Figure1 depicts the pipeline for testing. Input WSIs are processed and patches of 1024x1024 are used for stain generalization using CycleGAN. Patches of 512x512 are fed to RetinaNet for candidate detection of mitosis and mitosis-like structures. The predicted bounding boxes are used at 50x50 for classification as mitosis or non-mitosis with a confidence threshold of 0.7. The center point of candidates classified as mitosis is appended to the output.

\section{Results and Conclusion}
On the preliminary test set, the algorithm scores an F1 score of 0.52 with a precision of 0.71 and recall of 0.41 on "https://midog2021.grand-challenge.org". Figure \ref{fig:fig_cyc} depicts the effectiveness of CycleGAN in tackling stain variations across scanners. We wish to further investigate segmentation approaches in mitosis detection on stain generalization by CycleGAN.The code used for training the network will be made publicly
available on "https://gitlab.com/onwardassist/midog" repository after the final submission deadline.

\begin{figure}
    \centering
    \includegraphics[width=.7\linewidth]{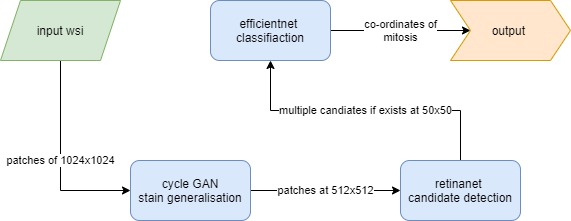}
    \caption{testing pipeline}
    \label{fig:testing_pipeline}
\end{figure}

\begin{figure}
\begin{subfigure}{.5\textwidth}
  \centering
  \includegraphics[width=.6\linewidth]{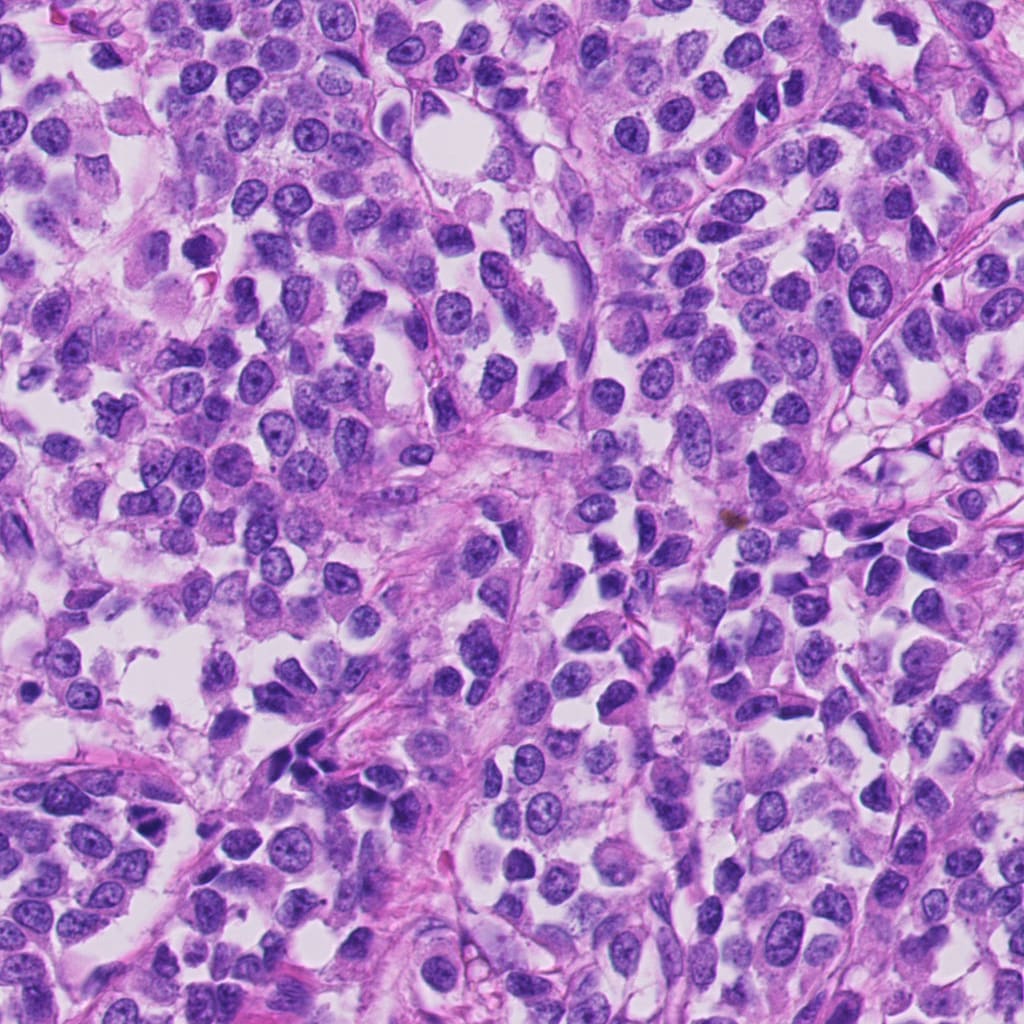}
  \caption{original patch from Hamamatsu XR NanoZoomer 2.0, }
  \label{fig:sfig1}
\end{subfigure}%
\begin{subfigure}{.5\textwidth}
  \centering
  \includegraphics[width=.6\linewidth]{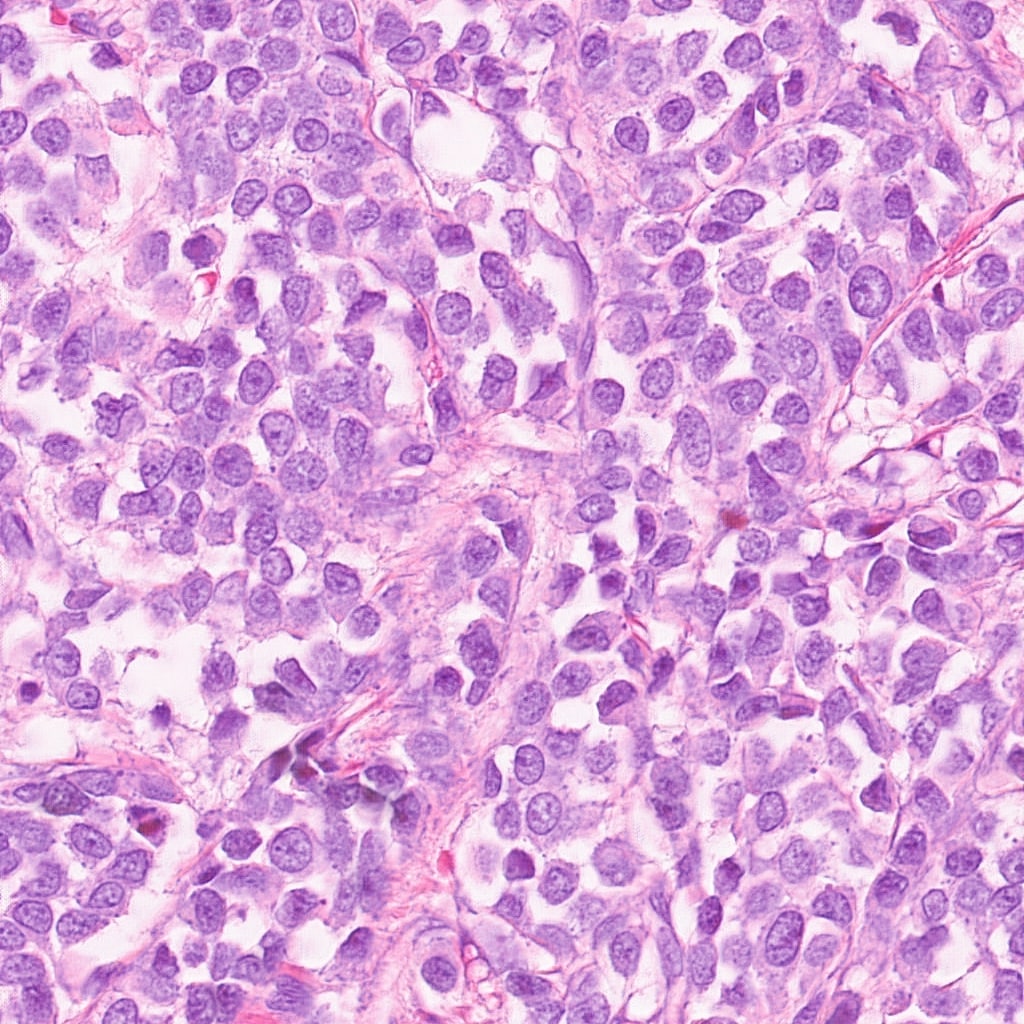}
  \caption{transformed patch for Hamamatsu XR NanoZoomer 2.0}
  \label{fig:sfig2}
\end{subfigure}

\begin{subfigure}{.5\textwidth}
  \centering
  \includegraphics[width=.6\linewidth]{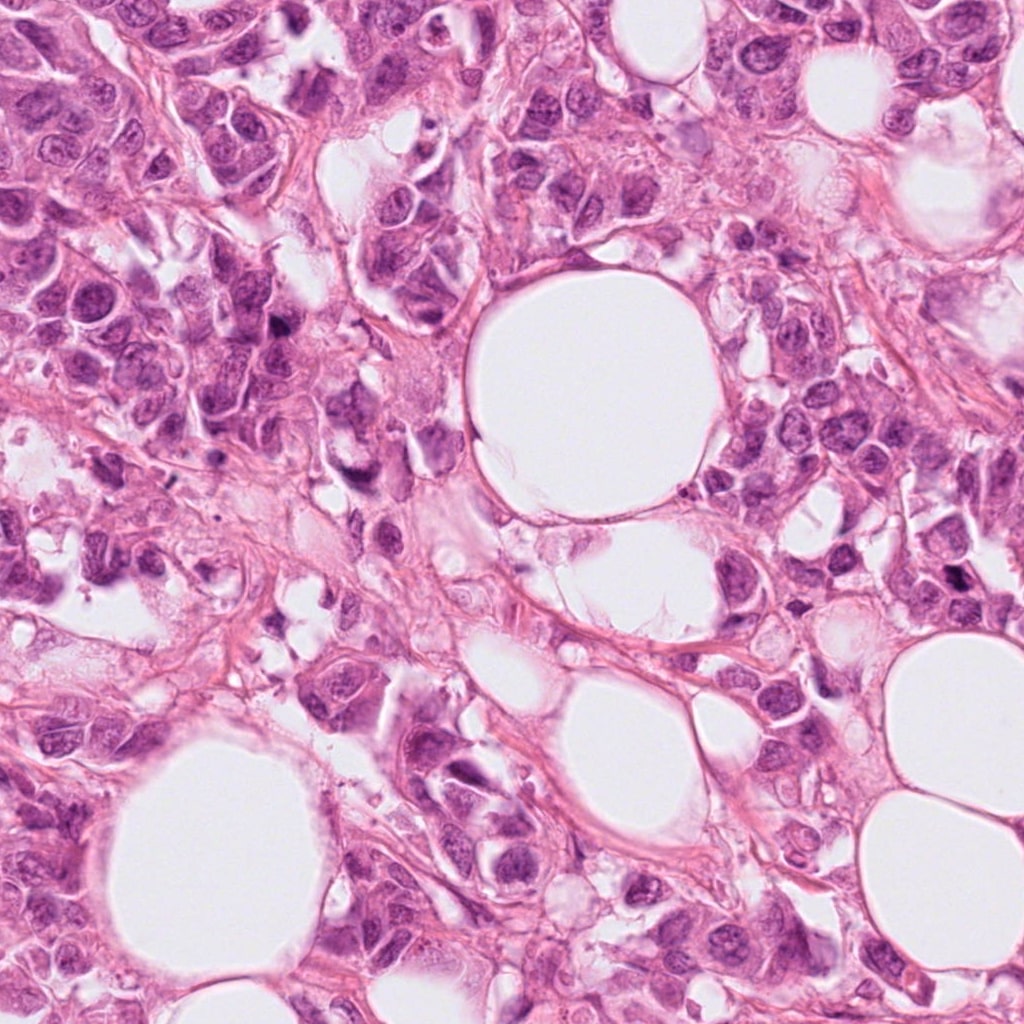}
  \caption{original patch from Hamamatsu S360}
  \label{fig:sfig5}
\end{subfigure}%
\begin{subfigure}{.5\textwidth}
  \centering
  \includegraphics[width=.6\linewidth]{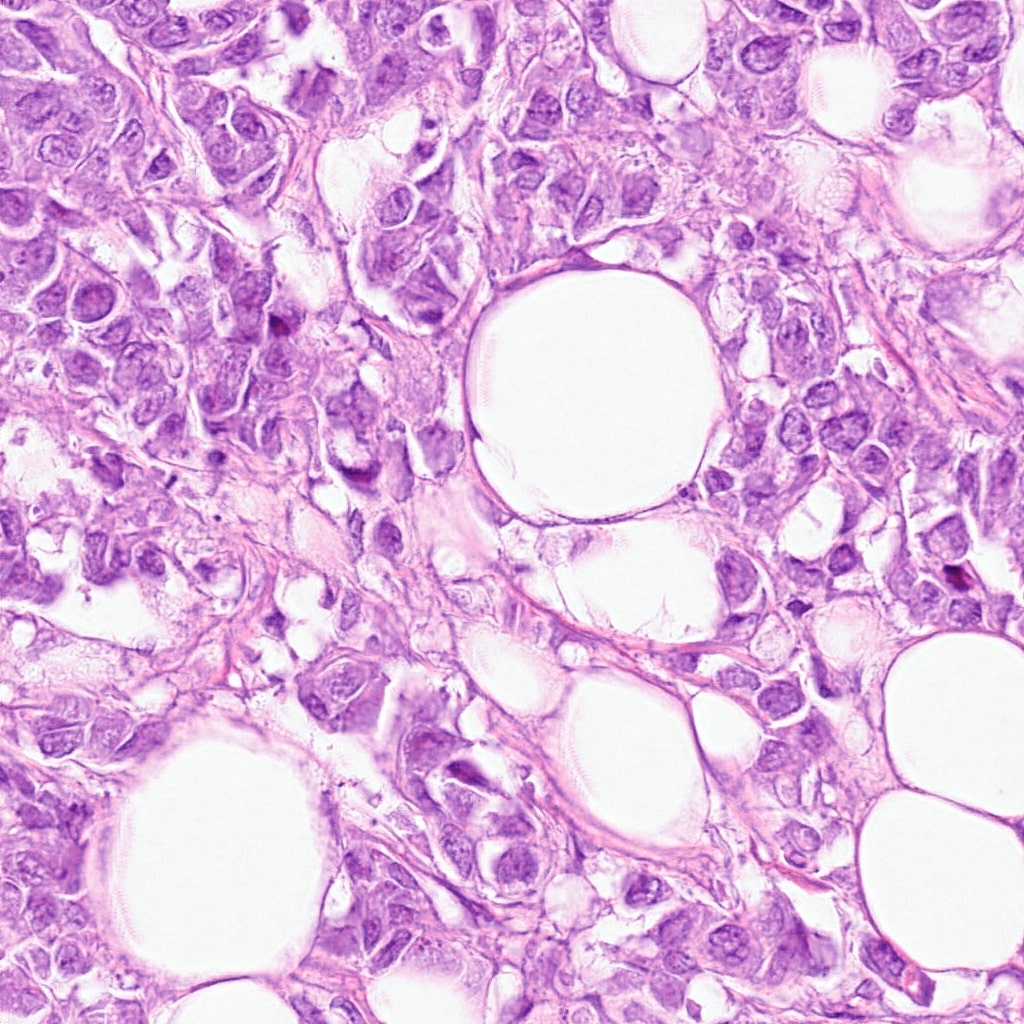}
  \caption{transformed patch for Hamamatsu S360}
  \label{fig:sfig6}
\end{subfigure}

\begin{subfigure}{.5\textwidth}
  \centering
  \includegraphics[width=.6\linewidth]{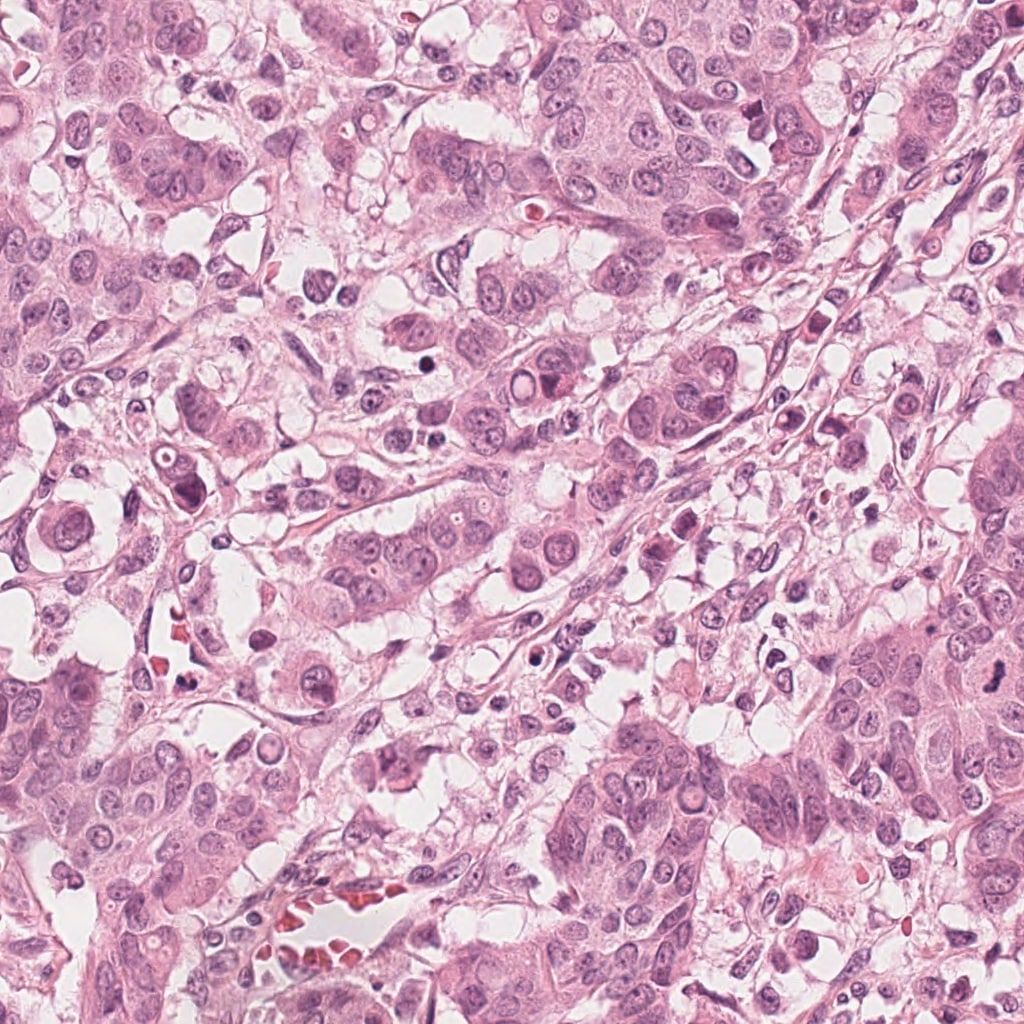}
  \caption{original patch from Aperio ScanScope CS2 }
  \label{fig:sfig7}
\end{subfigure}%
\begin{subfigure}{.5\textwidth}
  \centering
  \includegraphics[width=.6\linewidth]{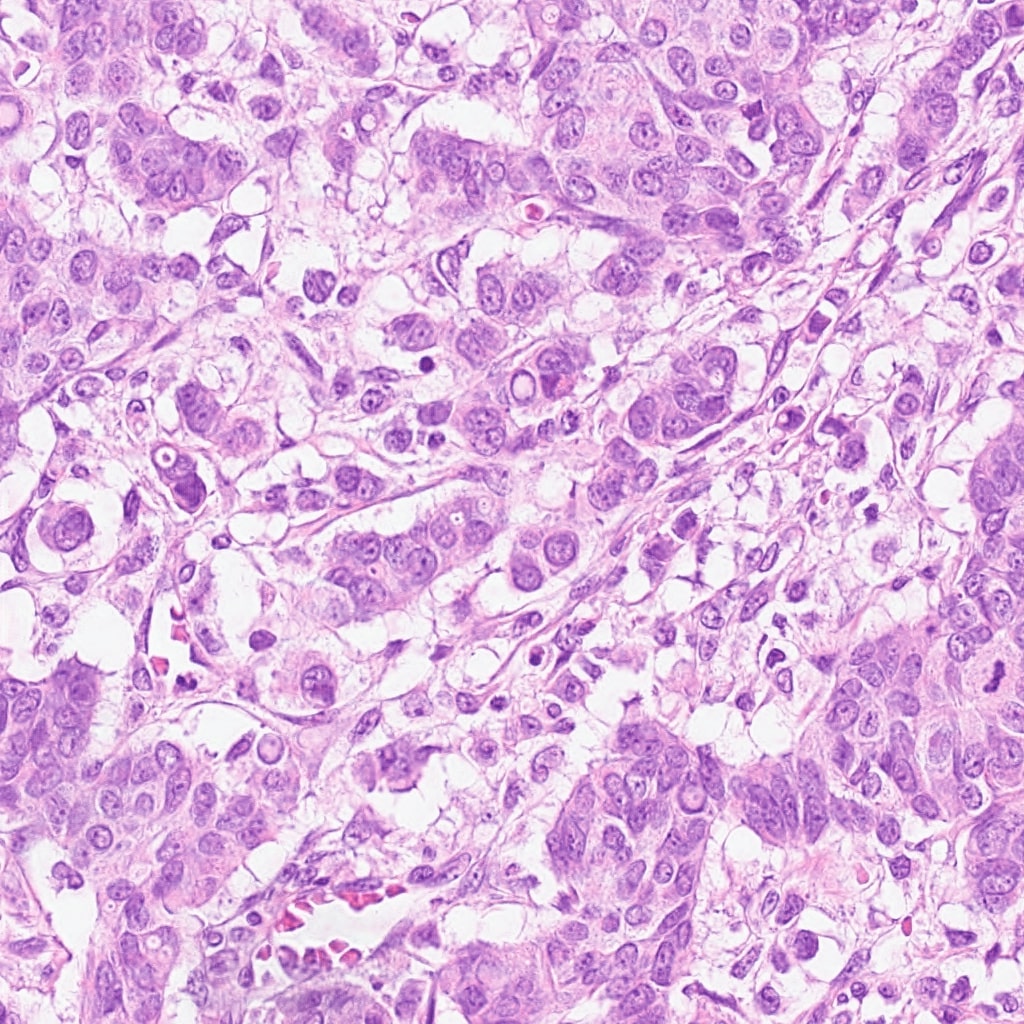}
  \caption{transformed patch for Aperio ScanScope CS2}
  \label{fig:sfig8}
\end{subfigure}

\begin{subfigure}{.5\textwidth}
  \centering
  \includegraphics[width=.6\linewidth]{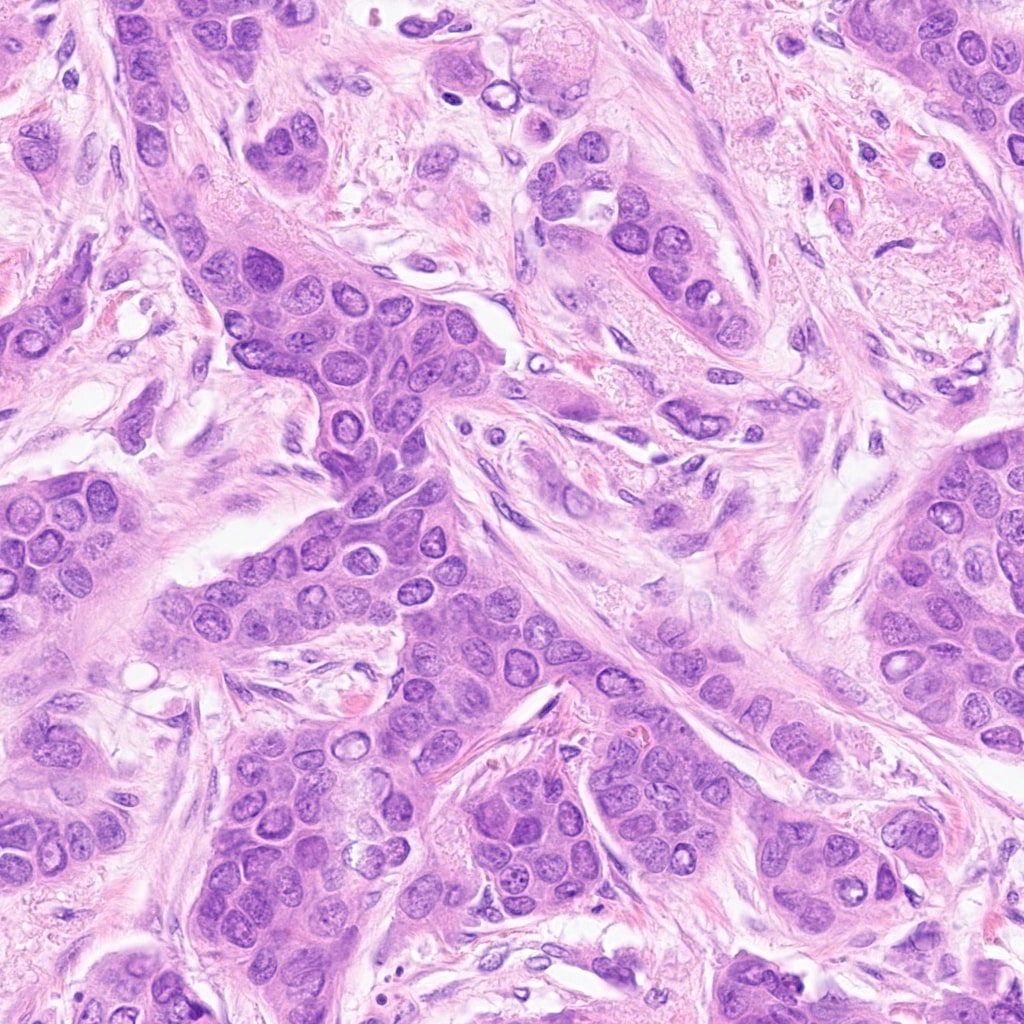}
  \caption{original patch from Leica GT450}
  \label{fig:sfig3}
\end{subfigure}%
\begin{subfigure}{.5\textwidth}
  \centering
  \includegraphics[width=.6\linewidth]{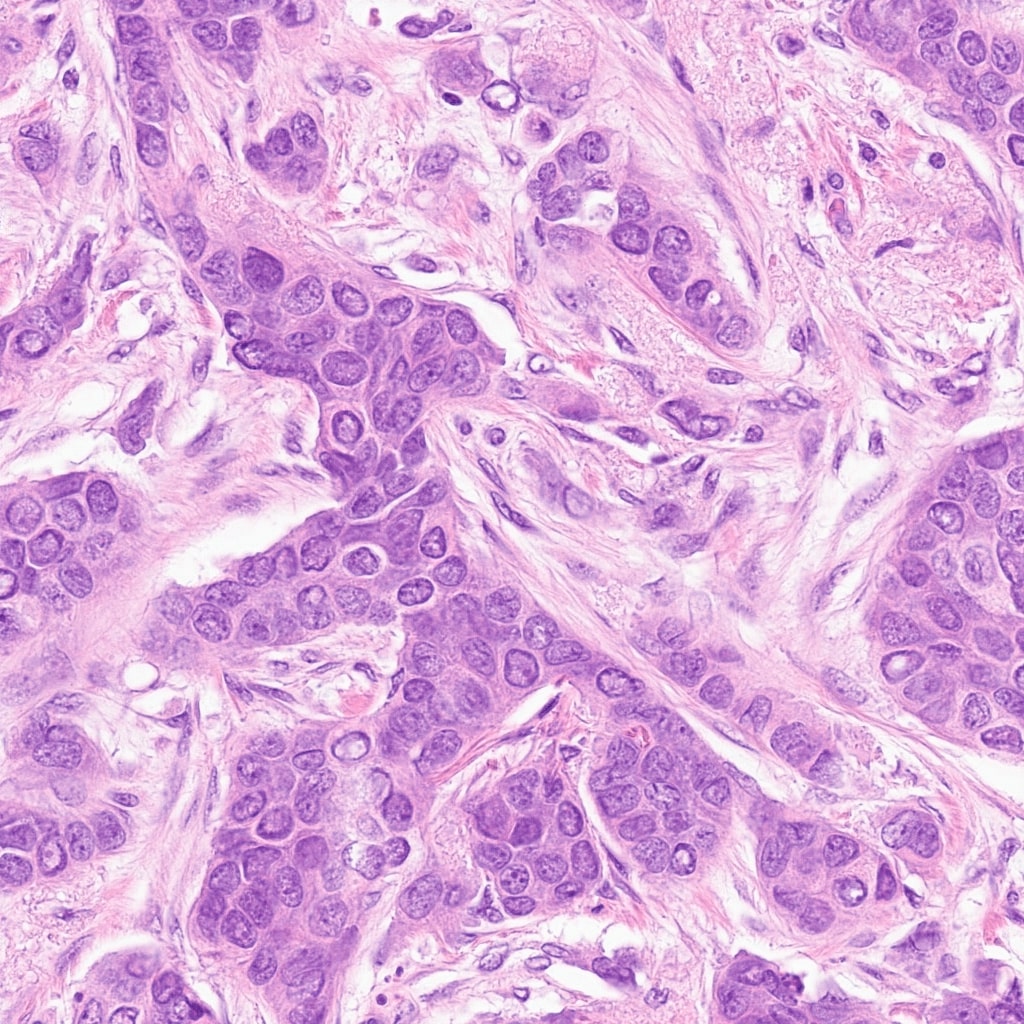}
  \caption{transformed patch for Leica GT450}
  \label{fig:sfig4}
\end{subfigure}

\caption{The plots show 1024x1024 patches from original and transformed patches. The domain variability is evident by visual differences in staining in (a), (c), (e), and (f). The transformed patches to the right (b), (d), (f), and (h) respectively shows consistent staining pattern while preserving morphology.}
\label{fig:fig_cyc}
\end{figure}

\bibliographystyle{unsrt}  
\bibliography{references}  






\end{document}